# Root causes, ongoing difficulties, proactive prevention techniques, and emerging trends of enterprise data breaches

**Miss. Gayatri Pise**
**Student, School Of Engineering**
**Ajeenkya DY Patil University**
**gayatri.pise@adypu.edu.in**

**Miss. Rina Patil**
**Student, School Of Engineering**
**Ajeenkya DY Patil University**
**Rina.patil@adypu.edu.in**

**Mr. Yatin Bhosale**
**Student, School Of Engineering**
**Ajeenkya DY Patil University**
**yatin.bhosale@adypu.edu.in**

**Co-Author**
**Dr. Ankita Agrawal**
**Faculty, School Of Engineering**
**Ajeenkya DY Patil University**
**ankitahctm@gmail.com**

A data breach in the modern digital era is the unintentional or intentional disclosure of private data to uninvited parties. Businesses now consider data to be a crucial asset, and any breach of this data can have dire repercussions, including harming a company's brand and resulting in losses. Enterprises now place a high premium on detecting and preventing data loss due to the growing amount of data and the increasing frequency of data breaches. Even with a great deal of research, protecting sensitive data is still a difficult task. This review attempts to highlight interesting prospects and offer insightful information to those who are interested in learning about the risks that businesses face from data leaks, current occurrences, state-of-the-art methods for detection and prevention, new difficulties, and possible solutions.

# INDEX

## INTRODUCTION

Organizations of all sizes, including governments and huge enterprises, are becoming increasingly concerned about data leaks. Sensitive information leaks, whether deliberate or unintentional, can have dire repercussions, including harm to an organization's brand and large financial losses. Numerous data kinds are at risk, including medical information, employee and customer data, and intellectual property. To put the financial damage into perspective, the average cost of a data breach has increased to $4 million, according to IBM's 2016 Cost of Data Breach Study. The rising digitization of our lives and corporate records is expected to result in a global yearly cost of data breaches exceeding $2.1 trillion by 2019, according to a forecast from Juniper Research.

Numerous high-profile data breaches that have cost businesses millions of dollars have occurred in recent years. For example, in 2013, hackers gained access to Target Corporation's network, stealing 70 million user identities and 40 million credit card details. Target has acknowledged losses of $248 million as a result of this event. A 2014 "state-sponsored" data breach that affected at least 500 million accounts was revealed by Yahoo in 2016.

With the exponential growth of data in the digital era and the increasing frequency of data breaches, protecting sensitive data from unauthorized access is an essential security concern for businesses.

There are several ways that data leaks can happen, including unintentional (like when employees or partners accidentally disclose it) and purposeful (such when data theft or insider sabotage). Remarkably, research has revealed that a sizable percentage of data breaches are the consequence of inadvertent activities, frequently as a result of breaches in security protocols or errors in corporate procedures. It's crucial to remember that insider threats—whether motivated by financial incentives, personal grudges, or corporate espionage—have a significant impact on data breaches.

Organizations use Data Leak Prevention and Detection (DLPD) systems in an effort to stop data leaks. These systems are designed to detect, track, and stop sensitive data from being accidentally or purposefully exposed. A variety of technical techniques are used by DLPD systems, including data traffic inspection, data usage policy enforcement, and modelling typical database access patterns.

Data leak prevention is becoming more and more difficult in the big data era. Overabundance of data exposes organizations to new threats even as it provides competitive advantages through enhanced business intelligence and personalized offerings. The proliferation of data and the widespread utilization of contemporary communication channels increase the number of possible avenues through which data may leak, including cloud file sharing, email, websites, instant messaging, and more.

The purpose of this review paper is to provide insight into the risks that data leaks present to businesses. It classifies various dangers, looks into well-known data breach instances, and summaries the lessons discovered. The study also explores methods created to identify and stop data breaches, highlighting the existing drawbacks of these strategies. In light of the difficulties associated with data leaks in the big data era, the paper presents a case study of a privacy-preserving data leak detection system and provides an overview of possible directions for future research in this important area.

### ENTERPRICES DATA LEAKS THREATS

Different taxonomies are available in the field of enterprise data security to categories hazards related to data leaks. We use these taxonomies in this part to classify and clarify important risks associated with data leaks. Furthermore, we examine actual data breach situations that occur in businesses and draw insightful conclusions from them that illuminate the lessons that may be applied.

These taxonomies are essential frameworks for comprehending the wide-ranging and dynamic field of hazards related to data leaks. We are able to get a complete picture of the difficulties that organizations encounter in protecting their sensitive data by looking at how various threats fall into these categories. Moreover, we may extract best practices and useful knowledge to improve business data security measures by doing a retrospective analysis of real data breach instances.

This section essentially outlines the main types of data leak threats and offers a practical viewpoint on the practical consequences of these concerns. For businesses trying to strengthen their defenses against a constantly changing range of threats in the digital era, this strategy is crucial.

## ANALYZATION OF RISK ACTIVITY FOR DATA LEAKS

There are various methods for categorizing the risks of data leaks, and each one throws light on the reasons and perpetrators of these security lapses. Whether these risks are the result of deliberate behavior or happen accidentally, one approach to classify them is according to their underlying causes. Another strategy separates insiders from outsiders based on who was responsible for the leak. Let's take a closer look at these divisions.

**Intentional vs. Inadvertent Leaks:**

One important distinction in the world of data leak dangers is the intentionality or inadvertency of the breach. Most intentional leaks are the result of malevolent insiders or outside parties. Techniques including malware, viruses, social engineering, and hacker infiltrations frequently lead to external data breaches. Cybercriminals, for example, may use system flaws or improperly set access controls to get around authentication procedures and obtain unauthorized access to private data. Social engineering techniques, including phishing attacks, have advanced in sophistication, deceiving workers into giving up important firm information to cybercriminals.

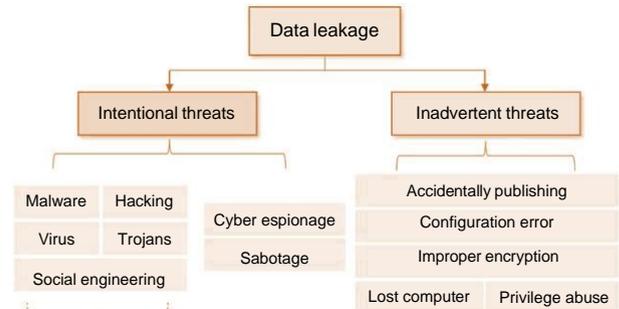

Figure 1st: categorization of risks related to corporate data leaks.

Conversely, intentional activities or inadvertent mistakes may result in internal data leaks. Corporate espionage may be the deliberate result of employee complaints or financial incentives. Conversely, inadvertent disclosures occur frequently when staff members share or send private information without using the appropriate encryption. Interestingly, studies show that insiders are responsible for more than 60% of data breaches, highlighting the importance of both technology and non-technological precautions in averting these kinds of incidents.

Other Classifications:

Other characteristics, including the industrial sector or the circumstances surrounding the incident, can also be used to categories data leaks. The Identity Theft Resource Centre (ITRC) has been tracking data that shows an increase in the overall number of significant data breach cases in recent years. The industries most frequently impacted have been business and medical/healthcare, with business data breaches accounting for 45.2% of all occurrences in 2016 and medical/healthcare accounting for 34.5%. An additional classification is based on the kind of occurrence; for example, "Other" includes categories such as staff errors and exposure to email or the internet. In this context, malevolent foreigners were responsible for over 55% of the breaches that occurred in 2016. Because different cybersecurity reports use different datasets, the results may differ slightly, but they all highlight a similar trend: insider threats, which now account for over 40% of all breaches, are the main cause of enterprise data leak threats.

To sum up, these categories offer a multifaceted perspective on the risks associated with data leaks, including information on the origins, offenders, industry-specific trends, and types of events. Comprehending these aspects is crucial for enterprises looking to strengthen their data security protocols in a digital environment that is changing quickly.

## INCIDENTS OF ENTERPRISE DATA LEAKS

| Organization | Records | Breach Date | Type | Source | Industry | Estimated Cost |
|---|---|---|---|---|---|---|
| Anthem insurance | 78 million | January 2015 | Identify theft | Malicious outsider | Healthcare | $100 million |
| Yahoo | 500 million | December 2014 | Account access | State sponsored[1] | Business | $350 million |
| Home depot | 109 million | September 2014 | Financial access | Malicious outsider | Business | $28 million |
| JPMorgan chase | 83 million | August 2014 | Identify theft | Malicious outsider | Financial | $13 billion |
| Benesse | 49 million | July 2014 | Identify theft | Malicious insider | Education | $138 million |
| Korea credit bureau | 104 million | January 2014 | Identify theft | Malicious insider | Financial | $100 million |
| Target | 110 million | November 2013 | Financial access | Malicious outsider | Business | $252 million |
| Adobe System | 152 Million | September 2013 | Financial access | Malicious outsider | Business | $714 Million |

TABLE 1st  Large-scale incidents of corporate data leaks in recent years

Sadly, we've seen an increase in the frequency of large-scale data breaches within businesses in recent years. Let's examine a few noteworthy examples from Table 1 that have all had far-reaching effects to highlight how serious the situation is. Numerous millions of people's personal information has been compromised by these breaches, which have also caused enormous financial losses—often amounting to hundreds of millions of dollars. To gain a deeper understanding of the impact of these breaches, we'll delve into a few recent incidents caused by external cyberattacks and those originating from insiders. In particular, we'll conduct an in-depth examination of the Target data breach, which stands as a representative example of an incident resulting from external attackers. These real-world examples highlight the importance for organizations to address the growing threat to their data security by effectively illustrating the enormous consequences that data breaches may have.

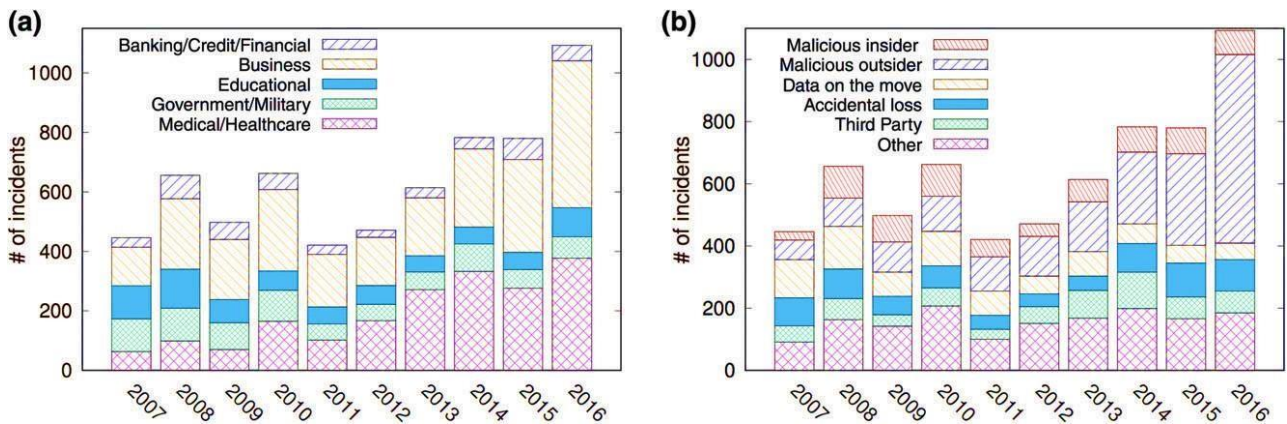

FIGURE 2nd   Recent data leak incident statistics include (a) industrial sector breaches and (b) type of breach data.

## Internal Data Leak Incidents

Sensitive information has inadvertently been made public in an alarming increase of unintended data leak instances in recent years. An example of this occurred in October 2016, when a worker for the Australian Red Cross Blood Service inadvertently uploaded papers on their website that contained the private information of over 550,000 blood donors in an insecure, publicly accessible directory. From 2010 to 2016, this sensitive data included names, addresses, dates of birth, and possibly sensitive information like medical histories, drug usage, and sexual activities. Similar to this, in 2011 a Texas server unintentionally released 3.5 million citizens' personal data online, where it was viewable for a full year. These illustrations show how often unintentional data exposure occurs, which emphasizes the necessity of strict data protection protocols. Insider acts that are purposeful can also be a threat, in addition to unintentional leaks. These include activities like eavesdropping, employing packet sniffing techniques, copying, extracting, and exfiltrating confidential data. In this instance, a notable data breach incident occurred in 2010 when more than 250,000 private documents from US diplomatic cables were made public. An internal person used an external hard drive to carry out this hack, which was eventually disclosed to WikiLeaks. About 100,000 diplomatic cables labelled "confidential" and 15,000 tagged "secret" were among the contents; they dealt with delicate political issues that attracted international interest.

In a different incident from 2013, an IT contractor working for Vodafone Germany obtained access to the company's database system and pilfered up to two million customers' bank account and personal information. Concerns were raised by this intrusion regarding possible phishing attempts aimed at the impacted clients. Vodafone responded to the issue by changing the administrator passwords and certificates, as well as completely wiping down the compromised server to stop any more data leaks.

Insider-caused data breaches have also affected the healthcare industry. The number of incidents of insiders mishandling private medical data has increased since medical records have gone digital. A former worker at UMass Memorial Medical Centre was charged in 2015 with stealing up to 14,000 patient details, which included Social Security numbers, names, dates of birth, and addresses. There's a chance that this breach lasted for twelve years.

Because people with legitimate access to organisational resources and data are frequently involved in these breaches, detecting internal data leak instances is a difficult process. Because of their extensive understanding of internal systems and strategies for getting around monitoring devices, their acts might not set off traditional alarms. Malicious insiders also have access to more advanced techniques, such as steganography and covert channels, which make it harder to detect their actions. For instance, they could use secret or encrypted channels to transfer confidential data that appears to be harmless.

Insiders have access to an increasing amount of sensitive data as we move through the big data era, which makes organisations' security concerns more pressing. To minimise inadvertent or unintentional data leaks, it is crucial to raise user knowledge about workplace security in addition to adopting technological measures.

## External Data Leak Incidents

Organisations have suffered enormous financial losses as a result of several high-profile data breach instances; Target and Ya hoo are two such examples. The 2016 Yahoo data breaches are notable for being among the biggest in history. In late 2014, hacker attacks resulted in the compromise of up to 500 million user accounts. Yahoo subsequently found evidence of yet another huge cyber intrusion in December 2016, which was thought to be unrelated to the first breach and had affected over 1 billion user accounts in August 2013. The repercussions were profound: Verizon paid $350 million less than the original agreed-upon sale price when acquiring Yahoo. Between November 27 and December 18, 2013, Target Corporation, one of the biggest stores in the country, experienced another serious breach. Target's data security was breached by cybercriminals, resulting in the theft of personal data belonging to up to 70 million consumers. This data included names, addresses, phone numbers, email addresses, and financial information.

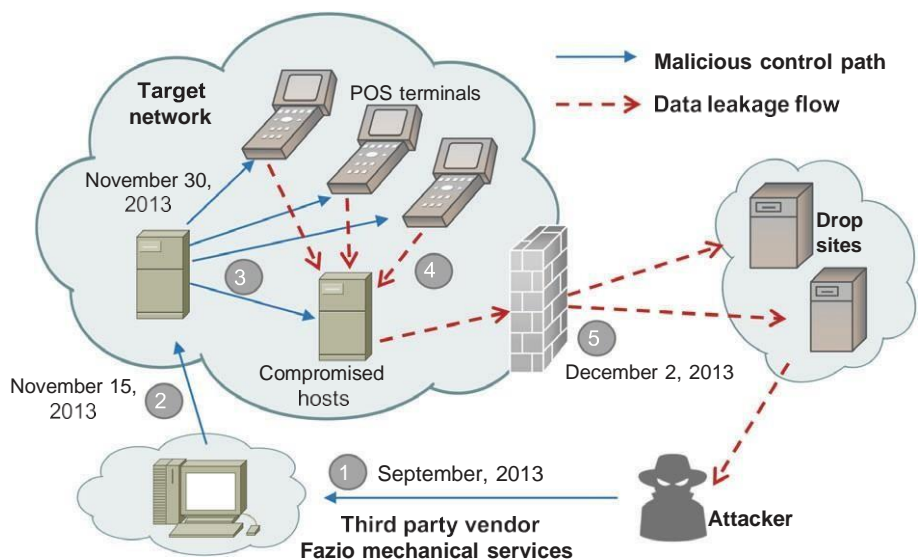

FIGURE 3 | Breakdown and analysis of the Target data breach.

Take a look at Figure 3 to get an idea of how the Target data breach happened. The attackers began their breach by using a phishing assault to gain access to the system of a third-party vendor, Fazio Mechanical Services (step 1). Fazio was able to remotely check store temperatures and energy usage thanks to his access capabilities to Target's network. Step 2: The attackers entered Target's network and found susceptible machines. They then broke into the point of sale (POS) networks and installed the data-stealing malware BlackPOS on the POS terminals (step 3). This malware has the ability to retrieve private data from POS devices' memory through scanning. The fourth step involves moving the encrypted stolen data from point-of-sale devices to hosts that had been infiltrated internally. Ultimately, in step 5, the hackers were able to effectively move the data to external drop sites, effectively removing it from the Target network. Adopting suitable technical and administrative methods for Data Leak Prevention and Detection (DLPD) is crucial, as the Target data breach provides as a grim reminder. The incident was caused by a number of significant technological errors in Target's security system, including: 1. Inadequate Access Controls: Target's failure to implement suitable access control measures on its partners in the third party allowed the first hacker infiltrations. 2. Network Segmentation: Vulnerabilities were created because the organisation failed to appropriately isolate the sensitive payment system from the other networks. 3. Weak POS Systems: Unauthorised software installation and configuration were made possible by Target's failure to harden the POS systems. 4. Disregarded Security Warnings: Target had network intrusion prevention systems and firewalls installed, but these products' security notifications went unanswered.

A key lesson for defenders is highlighted by this case: there are numerous ways to stop or identify data breaches. These consist of checking that code loads, limiting access for non-essential business partners, teaching staff members about the dangers of phishing, and using anomaly-based traffic monitoring to spot unusual destination and access patterns. Proactive security defences strategically deployed raise the bar for attackers and lower the likelihood of data leakage.

# DLPD TECHNIQUES

There is a wide range of approaches available in the field of Data Leakage Prevention and Detection (DLPD) techniques; the majority are from the research community, with only a small number provided by the industry. We'll examine the current DLPD methods and their drawbacks in this section.

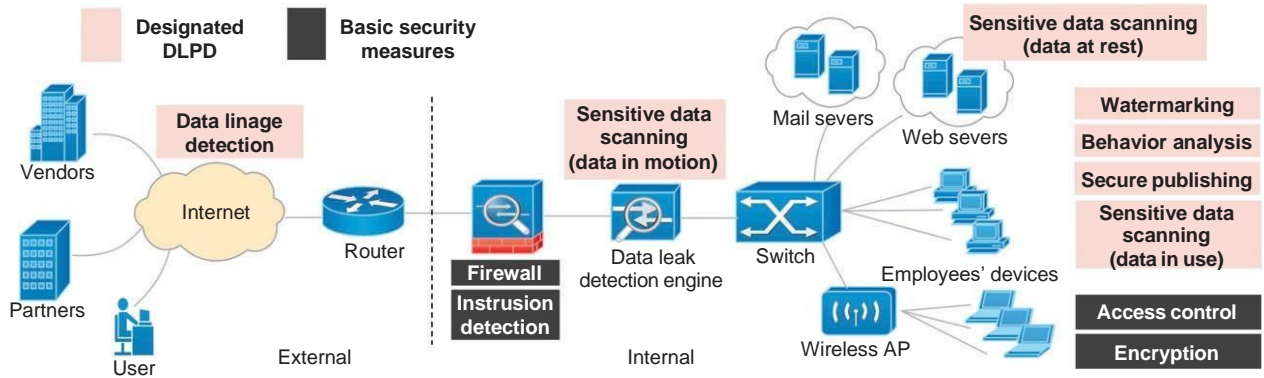

FIGURE 4ᵗʰ In an enterprise setting, there are several opportunities to use complementary data leak prevention and detection (DLPD) strategies.

## Categories of Existing DLPD Techniques:

DLPD techniques that are currently in use can be divided into two main groups: designated DLPD approaches and fundamental security measures. In contrast to standard security measures like as firewalls, antivirus programmes, intrusion detection, authentication, access control, and encryption, data leak prevention devices (DLPD) are specifically engineered to address the risks related to data leaks. Their primary duties include locating, keeping an eye on, and protecting private data from unwanted access. To find any data leaks, they usually use the monitored data's actual content or surrounding context. Designed DLPD products have grown more and more popular in the last few years, and they are now essential parts of business security packages.

Illustrating DLPD Techniques:
Typical methods for detecting and preventing data leaks are shown in Figure 4, along with an example of how they are implemented in an enterprise system. Data "at rest" is first protected by simple security techniques including encryption, secure data publishing, and limiting access to sensitive data. While intrusion detection systems (IDS) keep an eye on computer and network activity to spot unauthorised incursions, firewalls are responsible for restricting access to the internal network. Antivirus software helps detect malware that could be able to steal private data before it gets out, providing defence against intrusions from within.
It is important to remember that intrusion detection systems (IDS) have a tendency to identify hostile activity despite having a high false positive rate. Virtual machine technology is being used in emerging methods for safeguarding sensitive computer files, and trusted computing technology is being used to provide a hardware-based root of trust for content security.

Technological Means in DLPD:

The two primary types of technology tools used by DLPD are content-based analysis and context-based analysis.
1. Content-Based Analysis: This method entails examining the actual content of the data to stop sensitive information from being unintentionally exposed in any of its three states (in transit, usage, and rest). Although content scanning is quite good at preventing unintentional data loss, it can be easily circumvented by external or internal attackers, frequently by obfuscating data.
2.Context-Based Analysis: On the other hand, context-based techniques mainly concentrate on contextual analysis of the metadata linked to the data being monitored or the context in which the data is situated. Some hybrid approaches used in DLPD solutions analyse both context and content.

Balancing Content and Context:

material-based approaches typically provide higher detection accuracy than pure context-based analysis, even though the main objective of DLPD is to identify material as sensitive. Data scanning can be used at several phases to secure data, as Figure 4 illustrates. Businesses are able to determine the dangers of data leaks within their organisation by scanning data that is kept on servers when it is at rest. Sensitive data can be improperly handled and prevented from accessing the company network by keeping an eye on the data as it is being used. Furthermore, keeping an eye on network data streams while they are in transit helps stop private information from leaving or entering the company network.
To put it briefly, data security and privacy concerns are constantly changing, and while DLPD approaches are essential for protecting sensitive data, they also need to balance context and content analysis.

## Content-Based Approaches in DLPD:

Content-driven The core of DLPD is the search for known sensitive data that may be found on a variety of devices, including as servers, laptops, cloud storage, and even outgoing network traffic. These methods mostly rely on statistical analysis of the observed data, lexical content analysis or data fingerprinting.

- Data fingerprinting: To identify possible data leaks, this technique creates signatures or keywords for known sensitive content, which are then compared to the content under observation. Hash values or digests of a set of data can be used to create signatures. Researchers such as Shapira et al., for example, presented a fingerprinting technique that is robust to rephrasing by extracting signatures from the essential secret content and discarding irrelevant sections.
- Lexical Content Analysis: This method helps find sensitive data that adheres to clear patterns. For example, structured data such as credit card numbers, social security numbers, medical terminology, or geographic information can be found in documents using regular expressions. Users can generate custom signatures and rules with open-source network intrusion detection systems like Snort. These can then be matched with network traffic to detect attempts at data leaking.
- Statistical Analysis: In this case, the frequency of n-grams, or fixed-size sequences of consecutive bytes within a document, is the main focus of analysis. In statistical analysis, researchers also investigate item weighting systems and similarity measures, giving different relevance values to things (like n-grams) rather than considering them equally.

Collection intersection is a popular statistical analysis technique that compares two collections of shingles to find any sensitive information. The content sequences under observation and the sensitive data sequences that must remain inside the enterprise network are compared to determine the similarity score. The total of all shared item occurrence frequencies between the two collections, normalised by the smaller collection's size, is the intersection rate.

Machine Learning-Based Solutions:

Novel approaches to sensitive data detection have been made possible by recent developments in machine learning. Symantec utilises vector machine learning (VML) technology, for instance, to locate critical material inside unstructured data. This method gradually improves the precision and dependability of detecting sensitive information through training. Other academics have developed text categorization algorithms based on machine learning to automatically discern between enterprise documents that are sensitive and those that are not.
These machine learning techniques are useful tools in the continuous endeavour to properly protect sensitive information because they can adjust to changing data, even when it takes on a fuzzy or variable appearance.

## Context-Based Approaches in DLPD:

Context-based methods tackle the prevention and detection of data leaks from a new angle. They place more emphasis on identifying insiders or intruders by profiling users' typical behaviour than on the existence of sensitive data. These strategies are very helpful for reducing insider risks in businesses.
- User Conduct Analysis: A approach that simulates typical users' data access behaviours and sounds an alert when a user deviates from their established regular profile was proposed by Mathew and colleagues. This aids in reducing insider risks, particularly those that affect database systems. It's a method that considers how people generally obtain data and spots any variations from that pattern of behaviour.

- Detection of Abnormal Access Patterns: Bertino and colleagues presented a method for identifying unusual access patterns in relational databases. They accomplish this by searching through log files for database traces. When people with particular roles behave differently from the norm, it can be identified that they are role intruders in database systems using this method.
- - Identification of Insider Threats: A collection of algorithms and techniques for identifying malevolent insider behaviour was given by the senator and colleagues. They demonstrated how to recognise the subtle indicators of insider threats in the information systems of an organisation. This method uses a hybrid framework that combines anomaly- and signature-based methods to monitor user activity and identify unusual behaviour.
- Semantic Analysis: Gyrus developed a system to prevent malicious activities by capturing the semantics of user intent. This helps ensure that a system's behavior aligns with what the user intends. It's a way to prevent activities like manipulating a host machine to send sensitive data to unauthorized parties.
- Behavior Monitoring: Maloof's system monitors insider behavior and activity to detect malicious insiders who operate within their legitimate privileges but engage in activities outside the scope of their assignments.

Methodologies Based on Machine Learning:

A large number of these context-based strategies make use of machine learning and data mining methods. One benefit of machine learning in particular is that it can successfully identify outliers and deviations from normal behaviour without needing detailed explanations of unusual actions.

The lack of adequate training data presents a problem when using data mining or machine learning to find anomalies. This calls on the availability of a sufficient dataset to train the algorithm to identify deviations.

## Limitations of Current DLPD Approaches:

As opposed to conventional security measures, workable Data Loss Prevention and Detection (DLPD) systems must fulfil the following specifications:

- Selective Data Blocking: While permitting regular traffic, they ought to selectively block data flows carrying sensitive information.
- Defence Against Insider Threats: Data loss brought on by both naive staff and hostile insiders should be prevented by DLPD solutions.
- Preventing Data Exfiltration: DLPD systems should stop malware or attackers from stealing data from an organization's perimeter, even in the event that more conventional security measures are ineffective.

The difficulty of identifying and averting enterprise data breaches persists despite a great deal of study on DLPD.

Here are a few significant drawbacks of the methods used in DLPD today:

- Signature-Based Detection: This essential DLPD technique is based on fingerprint databases generated with common hash algorithms. It has wide coverage and is simple to install, but it can be circumvented if sensitive data is changed, which could result in false negatives. Furthermore, the considerable data indexing and comparison involved in processing vast amounts of content can be computationally demanding.
- Regular Expressions: To a certain extent, altered data leaks can be captured by DLPD systems that use regular expressions for exact and partial string matching. They can, however, lead to significant false positive rates and provide only a limited level of data protection.
- Collection Intersection: Sensitive information in unstructured textual data is usually found using this method. Although it has the potential to cause false alarms and has significant computational and storage costs, it can withstand certain alterations to the data.
- Behaviour Analysis: Understanding user intent and thwarting insider threats need the application of behaviour analysis. However, due to the temporal dynamics of context information, existing behaviour analysis-based methods may be prone to errors, leading to low detection rates and significant false positives.
- Scalability: Content-based DLPD techniques have trouble processing large amounts of content data quickly, which makes them unsuitable for use in the big data era.
- Watermarking: Watermarking has little practical use in DLPD due to its susceptibility to malicious removal or distortion.
- Honeypots: Because insiders might not interact with or use them, honeypot approaches have drawbacks.

- Encryption and Obfuscation: It is difficult to find data buried in the noise of typical outbound web traffic since existing DLPD algorithms are essentially unable to identify encrypted or obfuscated information leaks.
- Performance Evaluation: While industrial DLPD systems can stop unintentional data leaks, they might not be able to stop deliberate data breaches. Performance assessments have demonstrated that data obfuscation is a means by which viruses or insiders might evade security safeguards.

Although the present methods of data leak prevention (DLPD) are successful in stopping unintentional and plain-text breaches, they encounter difficulties in detecting encrypted or obfuscated information leaks, addressing insider risks, and guaranteeing scalability in the era of big data. In the realm of data loss prevention and detection, addressing these restrictions is still a challenge.

| Technique | Analysis | Pros | cons |
|-----------|----------|------|------|
| Fingerprinting | Content | Simple, Better coverage | Very sensitive to data modification |
| Regular expression | Content | Simple, Tolerate certain noises | Limited data protection, High false positive High computation and storage cost, Inapplicable toevolved or obfuscated data |
| Collection intersection[9] | Content | Wide data protection | Large training data, Complicated |
| Machine learning | Content/context | Resilient to data modifications,High accuracy | Large training data, High false positives |
| Behavior analysis | Context | Mitigate insider threats | Vulnerable to malicious removal or distortion |
| Watermarking | Context | Forensics analysis | Limited applications |
| Honeypots | context | Detect malicious insiders | Limited applications |

TABLE 2nd An Overview of Current DLPD Methods

## DLPD IN THE BIG DATA ERA:

Difficulties with Big Data Era Data Leakage Detection Systems:
Data leakage detection systems face several major issues in the big data era. Here, we address these issues and their implications for data security:

1. Increase in Leaking Channels: As business systems' data volumes continue to rise, so do the possible leakage channels. Customers and business partners are among the many stakeholders with whom sensitive data are frequently shared. External collaborations and cloud file sharing make the problem of data leakage even worse.
2.Human Factors: There is a higher chance of data leaks when an organization's mobile workforce works off-site. Determining the source of data leaks can be extremely difficult due to human factors, such as employees accidentally handling data improperly.
3. Encryption: Although crucial for data security, using encryption to safeguard data might make it more difficult to identify data leaks. When the data is encrypted, finding leaks in the data becomes difficult.
.4. Steganography: It is challenging to identify covert data breaches due to the technique of encrypting data inside other data. Malicious actors can conceal important information using this technique in files that appear innocent.

Big data presents a number of unique technical issues, including the following:

- Scalability: It is essential to be able to handle massive amounts of data, from megabytes to terabytes. Massive data volumes need to be handled by DLPD systems with efficiency, particularly in dispersed contexts like the cloud. Scalability facilitates early data leak detection and helps cut down on processing times for data.

- Privacy Preservation: It's critical to protect sensitive data from prying eyes, including the DLPD provider and any hackers who might compromise the detection system. Maintaining privacy is a top priority, particularly when using outside providers to detect data leaks.
- Accuracy: For effective detection, low false negative and false positive rates are necessary. Accurate leak detection is challenging in big data environments because of their distributed nature. When data is contracted out to outside vendors, it's possible for it to be transformed or altered. For example, formatting tags or metadata may be added, or characters may be changed for style. The accuracy of content-based detection techniques may be impacted by these modifications.
- Timeliness: It's critical to have the capacity to identify and address data breaches almost immediately. Big data environments, which are defined by high data volume, diversity, and velocity, present benefits as well as difficulties in quickly recognising potential data breach concerns.

In conclusion, scalability, privacy preservation, accuracy, and timeliness are some of the challenging challenges that data leakage detection systems face in the big data era. In a digital world that is changing quickly, protecting sensitive information requires addressing these issues.

## A CASE STUDY:

We offer MR-DLD, a case study of a privacy-preserving data leak detection system, in response to the issues raised earlier. MR-DLD makes use of the MapReduce distributed computing architecture, which is widely employed in data-intensive applications like traffic/log analysis and spam filtering. The system can be implemented on local computer clusters or in cloud environments, with its main objective being to examine sensitive content for unintentional data leaks.

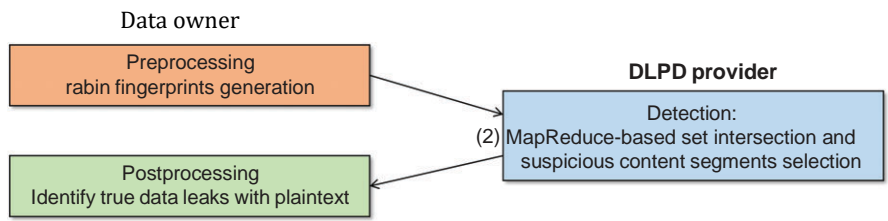

FIGURE 5th Distribution of workload between the provider of data leak prevention and detection (DLPD) and the data owner..

Crucial Elements of MR-DLD:

1. MapReduce Framework: To identify sensitive data patterns in massive amounts of information that are transferred over networks or kept in data storage, MR-DLD uses the MapReduce distributed computing framework. MapReduce is a good fit for the problem since it allows distributed data processing.

2. Privacy Preservation: Data owners may contract with outside DLPD providers to handle privacy-related data leak detection responsibilities. These providers, on the other hand, could be regarded as "honest-but-curious," which means that while they adhere to the guidelines, they might try to reconstruct sensitive material. Data owners change their data before to delivering it to MapReduce nodes in order to reduce the possibility of sensitive material being exposed during scanning. Rabin fingerprints, or one-way hash values of n-grams, are used by MR-DLD to offer robust yet effective confidentiality protection for sensitive content.

Workload Distribution:

burden Distribution: DLPD providers and data owners share the burden in MR-DLD. The DLPD supplier receives sensitive and content fingerprint collections from data owners. Then, using two-phase MapReduce algorithms—map and reduce operations—the supplier implements the MapReduce framework to compare these collections. Sensitive data leaks can be found using MR-DLD by calculating the intersection rate for each pair of content and sensitive collection. Based on predetermined thresholds, it notifies the data owner of any data leak warnings. Data owners get alerts, retransform them into unencrypted sensitive sequences and suspicious content chunks, and find where leaks happen.

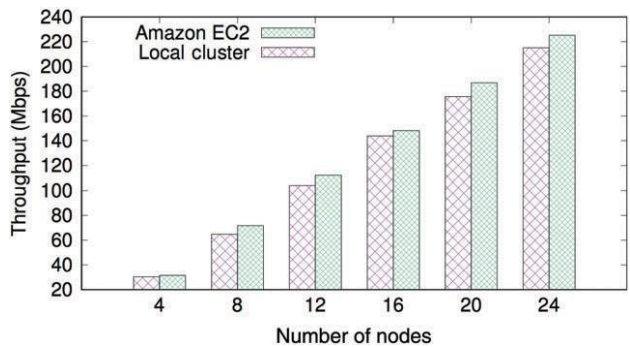

Evaluation of Scalability:

MR-DLD was tested with four to twenty-four nodes and different content amounts in order to determine its scalability. We ran experiments on Amazon EC2 as well as local clusters. The outcomes showed that when the number of nodes rises, the system grows successfully. On the local cluster, the highest throughput was 215 Mbps, while on Amazon EC2, it was 225 Mbps. A 3–11% increase in performance was seen by the EC2 cluster, in part because of its increased memory capacity.

To summarise, MR-DLD provides an efficient method of detecting data breaches in big data environments by using the MapReduce framework and giving priority to accuracy, scalability, and privacy preservation. It provides a workable answer to the problems brought on by the big data era of big data

## FURTHER RESEARCH OPPORTUNITIES:

Prospects for Future Research on Data Leakage Prevention and Detection (DLPD):

1. Using Deep Learning to Identify Insider Threats: Data mining and machine learning will be essential components of DLPD in the big data space. In a variety of applications, deep learning methods such as Deep Neural Networks have demonstrated potential in anomaly detection. These methods can be used for DLPD content and context analyses, which will allow for the prompt preservation of sensitive data while improving accuracy and identifying minor data leaks. Additionally, deep learning may help close the semantic gap that exists between low-level machine events and high-level user intentions. Because user intentions are important yet difficult to quantify directly, there is a gap that is frequently seen in insider threat detection. Sequences of machine events may be used by deep learning models to learn to infer user intents, opening up exciting new research opportunities.

2. DLPD as a Cloud Service: The emergence of cloud computing opens up new avenues for data leak detection research. Businesses can contract with outside companies to handle data processing, but this creates privacy issues. If sensitive data is outsourced to a third party with enough background frequency information, the collection intersection approach—which depends on the similarity of two sets based on element frequency information—may be susceptible to frequency analysis. Algorithms for privacy-preserving data leak detection that are resistant to powerful attacks are needed. One important area of inquiry is how cloud service providers may scale without compromising detection accuracy or experiencing considerable latency while handling massive datasets. In this situation, streaming data handling frameworks such as Spark and Flink may be quite important.

3. Tracking Encrypted Channels: Current DLPD techniques frequently have trouble processing material that has been heavily modified, obscured, or encrypted. One of the biggest obstacles to content-based detection is encrypted traffic. Effectively monitoring encrypted connections to identify covert data leaks is a critical issue. Using differential analysis or data flow tracking are two such approaches. For instance, academics have looked into differential analysis methods for detecting privacy leaks in smartphone systems that are resistant to obfuscation. Future DLPD efforts to detect the transfer of sensitive information across encrypted channels may find use for research in the rapidly developing field of string matching on encrypted data.

4. Tracking Encrypted Channels: Current DLPD techniques frequently have trouble processing material that has been heavily modified, obscured, or encrypted. One of the biggest obstacles to content-based detection is encrypted traffic. Effectively monitoring encrypted connections to identify covert data leaks is a critical issue. Using differential analysis or data flow tracking are two such approaches. For instance, academics have looked into differential analysis methods for detecting privacy leaks in smartphone systems that are resistant to obfuscation. Future DLPD efforts to detect the transfer of sensitive information across encrypted channels may find use for research in the rapidly developing field of string matching on encrypted data.

## CONCLUSION

Organisations must devote all of their resources and persistent effort to stopping and identifying data leaks. We have thoroughly reviewed data leak dangers and essential methods for Data Leakage Prevention and Detection (DLPD) in this study. Although surveys that already exist provide thorough explanations of these methodologies, this review article has made an effort to highlight the ongoing difficulties, especially in the big data era. In addition, we have identified a number of promising research directions that may help reduce the risk of data breaches in business environments.

Based on our investigation, we have determined that two research topics are most promising:

1. Cloud-Based Data Leak Detection Services: The idea of contracting out data leak detection to outside companies is appealing, especially in light of the development of cloud computing. Concerns about data confidentiality and privacy are brought up by this change. Researchers want to concentrate on creating privacy-preserving algorithms and solutions that can safeguard private information while taking advantage of the cloud's scalability benefits.

2. Anomaly Detection for Insider Threats Using Deep Learning: Machine learning techniques are becoming more and more important in the field of big data, which generates large volumes of heterogeneous data. In particular, deep learning has a lot of promise for identifying insider threats, enhancing leak detection precision, and guaranteeing prompt defence. This approach may be able to close the semantic gap—a major problem in insider threat detection—between high-level user intentions and low-level machine occurrences.

These research avenues present worthwhile chances to improve Data Leakage Prevention and Detection and strengthen data protection methods in enterprise environments, as organisations confront constantly changing data security concerns.